\documentclass[prx,twocolumn,preprintnumbers,amsmath,amssymb,letterpaper]{revtex4}

\usepackage{graphicx}
\usepackage{latexsym}
\usepackage{amsmath}
\usepackage{graphics}
\usepackage{amssymb}
\usepackage{layout}
\usepackage{verbatim}
\usepackage{amsfonts,epsfig}
\usepackage{slashed}
\usepackage{feynmp}
\usepackage{float}
\usepackage{color}

\newcommand{\beq}{\begin{equation}}
\newcommand{\eeq}{\end{equation}}
\newcommand{\bea}{\begin{eqnarray}}
\newcommand{\eea}{\end{eqnarray}}

\begin{document}

\title{Deterministic generation of all-photonic quantum repeaters from solid-state emitters} 

\author{Donovan Buterakos, Edwin Barnes, and Sophia E. Economou\footnote{economou@vt.edu}}

\affiliation{Department of Physics, Virginia Tech, Blacksburg, Virginia 24061, USA}

\begin{abstract}
  Quantum repeaters are nodes in a quantum communication network that allow reliable transmission of entanglement over large distances. It was recently shown that highly entangled photons in so-called graph states can be used for all-photonic quantum repeaters, which require substantially fewer resources compared to atomic-memory based repeaters. However, standard approaches to building multi-photon entangled states through pairwise probabilistic entanglement generation severely limit the size of the state that can be created. Here, we present a protocol for the deterministic generation of large photonic repeater states using quantum emitters such as semiconductor quantum dots and defect centers in solids. We show that arbitrarily large repeater states can be generated using only one emitter coupled to a single qubit, potentially reducing the necessary number of photon sources by many orders of magnitude. Our protocol includes a built-in redundancy which makes it resilient to photon loss.
\end{abstract}

\maketitle

\section*{Introduction}

Quantum entanglement is the cornerstone of novel quantum technologies, particularly quantum computing and communication. The enormous interest in quantum communication is driven by its built-in security which protects the transmission of information and quantum entanglement for applications such as distributed quantum computing, quantum key distribution and quantum internet.  To create distributed entanglement between two separated locations (nodes), qubits at each node must be entangled to flying qubits which are then sent to a common location to undergo joint measurements that implement entanglement swapping \cite{Bouwmeester_Nature97,Hensen_Nature15}. To achieve entanglement between nodes at large distances, beyond the capabilities of optical fibers, the entanglement must be refreshed at intermediate nodes known as quantum repeaters \cite{Briegel_PRL98,Dur_PRA99,Pirandola_arxiv15,Wallnofer_PRA16}.
The standard paradigm for a quantum repeater is based on atomic quantum memories located at primary nodes, each entangled with a single photon \cite{Togan_Nature10}, which in turn is sent to a secondary, intermediate node \cite{Duan_Nature01}. Two photons arriving at a secondary node undergo a Bell measurement, a process that transforms the qubit-photon entanglement into long-distance two-qubit entanglement where the two qubits involved are located at the two different primary nodes on either side of the secondary node. Difficulties associated with atomic-memory-based quantum repeaters include the necessity for long coherence times of the atomic memory beyond what is currently feasible and also the inherent vulnerability of these schemes to photon loss.

\begin{figure}
	\centering
	\includegraphics[width=\columnwidth]{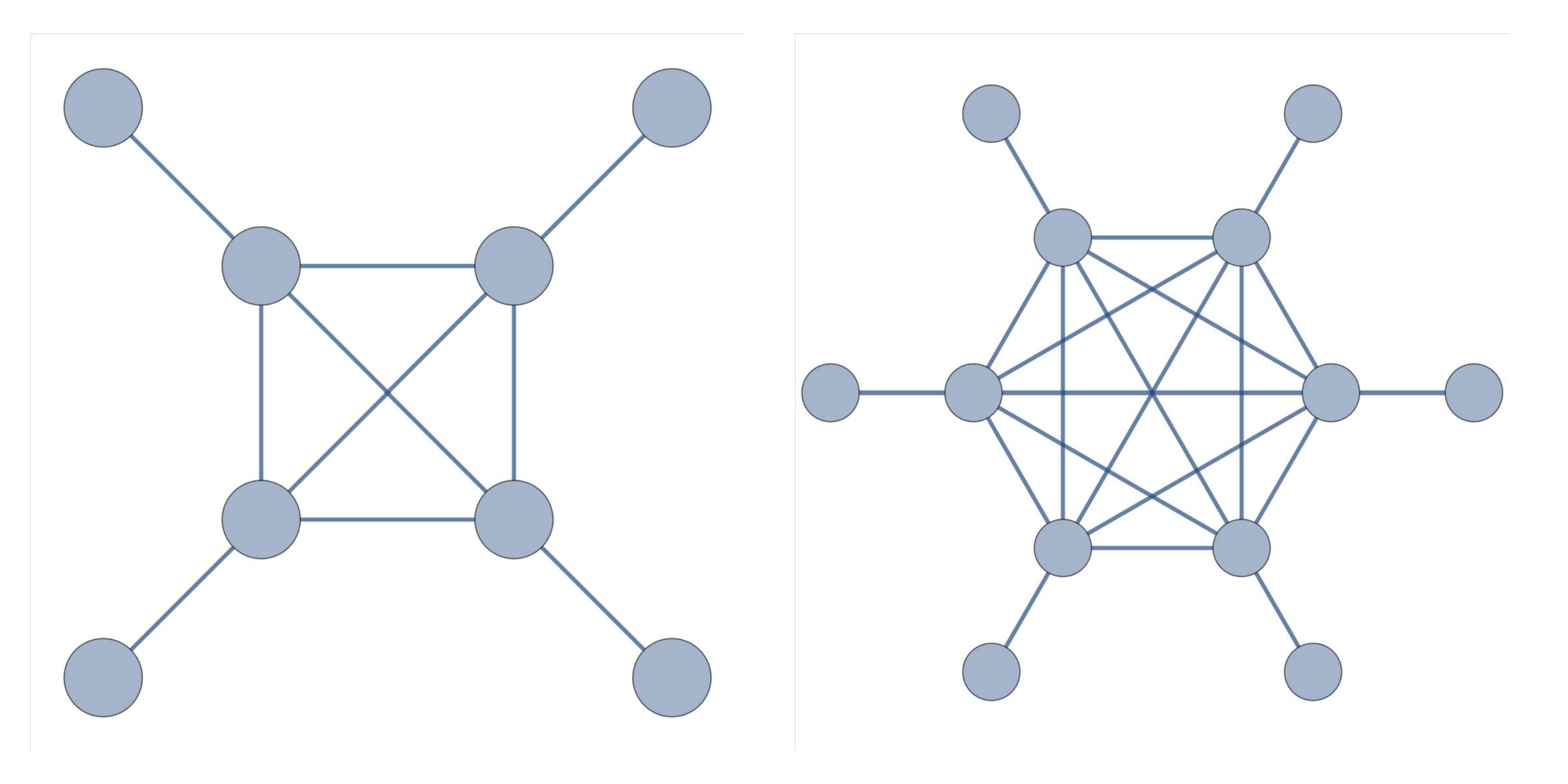}
	\caption{Graphical representation of $N{=}4$ and $N{=}6$ repeater states proposed in Ref.~\cite{Azuma_NC15}. States consist of a complete subgraph of $N$ core photons each connected to an additional photon forming N external arms.}
	\label{fig:rgs}
\end{figure}

Zwerger et al. \cite{Zwerger_PRA12,Zwerger_PRL13} introduced the idea of using highly entangled states, known as graph states \cite{Briegel_PRL01,Raussendorf_PRL01,Raussendorf_PRA03,Hein_PRA04,Nielsen_PRA05}, to implement quantum repeaters. In a recent publication, Azuma et al. \cite{Azuma_NC15} put forward an explicit construction of an all-photonic repeater graph state (RGS) consisting of a completely connected graph of ‘core’ photons, with each of them featuring a connection to an additional ‘arm’ photon, to be used for entanglement swapping in the secondary nodes. Two such states are illustrated in Fig.~\ref{fig:rgs}. In this figure, each circle represents a photon, and the lines between them represent pairwise entanglement \cite{Raussendorf_PRA03,Hein_PRA04,Nielsen_PRA05}. Specifically, the state represented by a graph can be created by initializing all qubits to the state $|+\rangle=(|0\rangle+|1\rangle)/\sqrt{2}$ and successively performing CZ gates between all pairs of qubits connected by an edge of the graph. This work has attracted a great deal of attention \cite{Takeoka_NC14,Bruschi_PRA,Azuma_NC15b,Pant_PRA17,Zwerger_APB16} due to its advantages over atomic-memory based repeaters, notably the all-photonic construction that avoids coherence time limitations, the resilience against photon loss, and the elimination of long-distance heralding \cite{Azuma_NC15}. These attractive features position quantum repeaters, and consequently long-distance quantum communication, as near-term, much more readily feasible technology compared to quantum computing \cite{Sinclair_PRL14,Azuma_NC15b}.

Despite the promise held by the all-photonic RGSs of Ref.~\cite{Azuma_NC15}, construction of multiphotonic entanglement is extremely challenging. This is due to the fact that photons do not interact with each other, so that either a nonlinear interaction or measurement are required to entangle two photons that are initially in a product state. Both approaches are challenging because non-linear interactions are weak and measurements are probabilistic. The standard process of generating graph states (or cluster states \cite{Raussendorf_PRL01} for quantum computing)\cite{Nielsen_PRL04,Nielsen_PRA05,Zwerger_APB16} begins with pairs of photons that are prepared in entangled (Bell) states through parametric down-conversion. Two different pairs are then ``fused" together probabilistically via the joint measurement of two photons, one from each pair. When the measurement succeeds, which happens with probability $1/2$ (or $3/4$ if ancillary qubits are used \cite{Grice_PRA11}), a three-photon graph state is obtained. Such fusion gates are used consecutively to `grow' the graph state \cite{Browne_PRL05,Varnava_PRL08}. To date, up to ten photons have been entangled in this way \cite{Zhao_Nature04,Gao_PRL10,Gao_NP10,Wang_PRL16b}. The analysis of quantifying the resource overhead of RGSs was carried out \cite{Pant_PRA17} and, even after an optimization of the original scheme of Ref.~\cite{Azuma_NC15}, $10^6$ photon sources per node are needed. Overcoming the probabilistic nature of RGS generation would therefore be catalytic in drastically reducing overhead.
Here, we show that repeater graph states of arbitrarily large size can be created deterministically by using quantum emitters with appropriate level structures and selection rules. The overhead of our approach is dramatically reduced compared to fusion-based approaches. Our remarkable finding is that the number of emitters required does not scale with the size of the repeater state: surprisingly, one emitter suffices to generate an RGS of arbitrary size.

\begin{figure}
	\centering
	\includegraphics[width=\columnwidth]{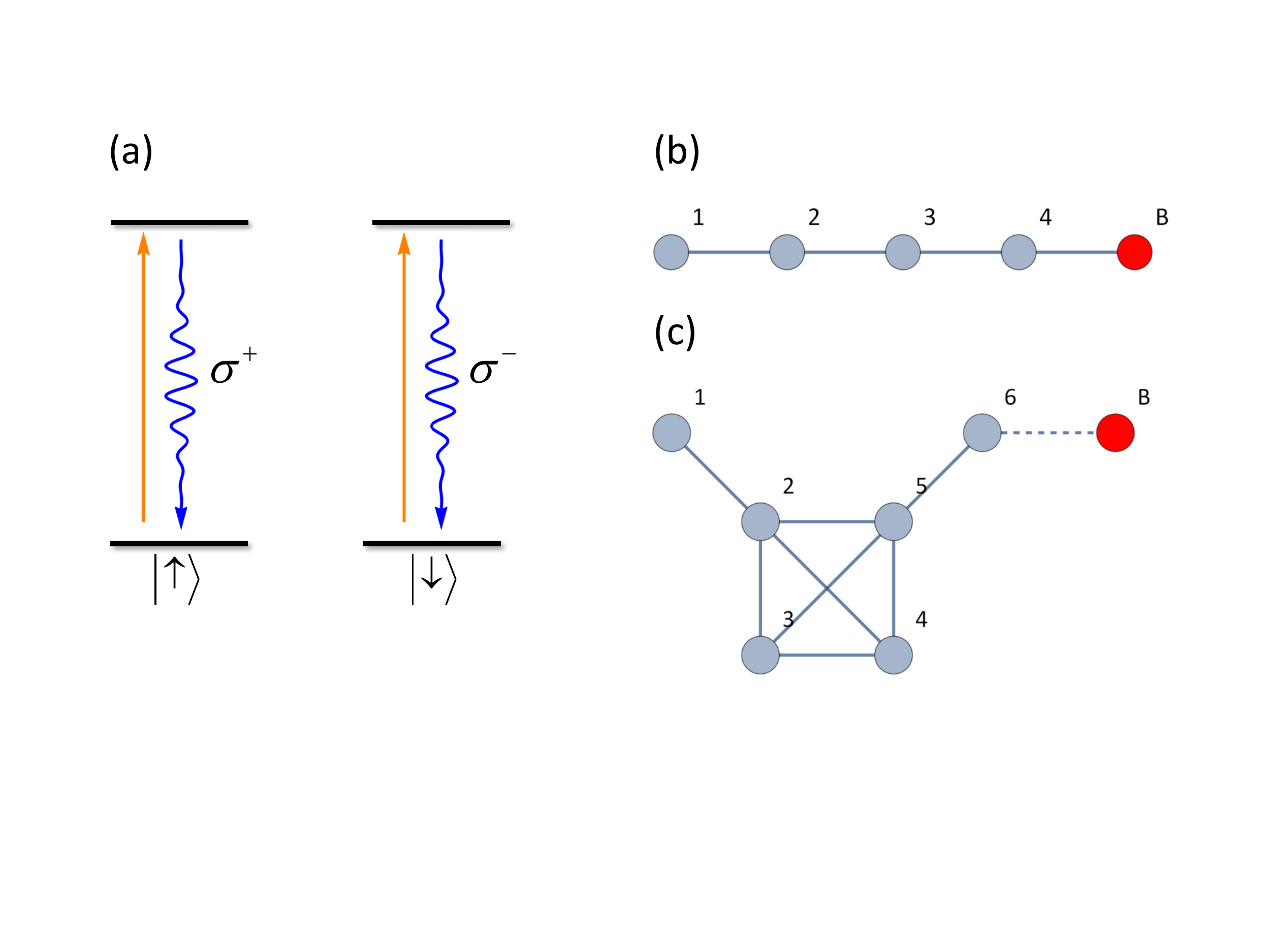}
	\vspace{-2cm}
	\caption{(a) Level structure required for the emitter to emit entangled photons. Each of two ground states couples to its own excited state. Cross transitions are forbidden by selection rules. (b) Linear photonic cluster state. Photons are blue circles, while the red circle is the emitter. Each line represents entanglement between the qubits it connects. (c) Partial $N{=}4$ repeater state. All but 2 of the 8 entangled photons comprising this repeater state can be produced from a single emitter.}
	\label{fig:energylevels}
\end{figure}

A deterministic approach to the creation of linear cluster states was introduced several years ago \cite{Lindner_PRL09}, and later generalized to a cluster state ladder \cite{Economou_PRL10}. The first experimental demonstration of deterministic linear cluster state generation was carried out recently following these ideas \cite{Schwartz_Science16}. The number of emitters needed to create larger two-dimensional cluster states is equal to the linear size of the state, making the creation of a large square grid cluster state conditional on future advances in controllably coupling a long chain of emitters. Nevertheless, the concepts introduced in Refs.~\cite{Lindner_PRL09,Economou_PRL10} for one and two emitters are central to our present work. We follow these works in assuming an emitter that has the level structure shown in Fig.~\ref{fig:energylevels}(a). We also consider that two adjacent emitters can be coupled such that entanglement can be created between them.


\section*{Results}

Lindner and Rudolph \cite{Lindner_PRL09} proposed a method in which a quantum dot or similar system with the level structure displayed in Fig.~\ref{fig:energylevels}(a) could be optically pumped to generate photons that are entangled with the electron spin. Moreover, they showed that repeated pumping of such an emitter can produce a chain of photons which can be entangled with the emitter and with each other. In particular, if the pumping is performed repeatedly without applying any other operations on the emitter, then a photonic GHZ state will be created in which every photon is entangled with the emitter and with every other photon. On the other hand, if a Hadamard gate is applied on the emitter between each pumping operation, then the resulting state is instead a one-dimensional linear cluster state in which each photon is entangled with the photon that preceded it and with the one that follows it. Fig.~\ref{fig:energylevels}(b) shows the graph corresponding to this linear photonic cluster state. 

\begin{figure*}
	\centering
	\includegraphics[width=2\columnwidth]{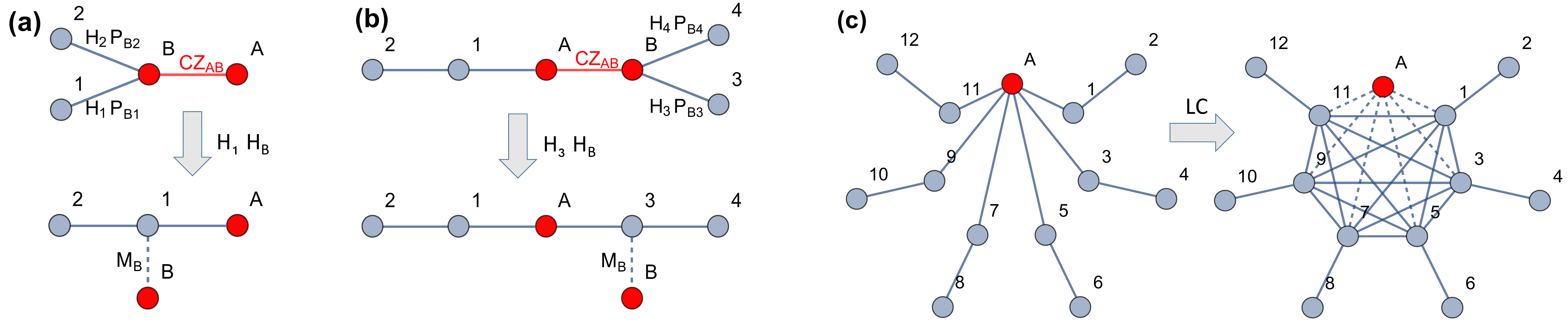}
	\caption{Generation of the $N{=}6$ repeater state proposed in Ref.~\cite{Azuma_NC15}: \textbf{(a)} Top: Emitter $B$ is connected to ancilla $A$ with a CZ gate and pumped twice, creating photons 1 and 2. Bottom: Hadamard gates are applied and emitter $B$ is measured, detaching it from the graph. \textbf{(b)} This process is repeated (creating photons 3 and 4), generating another arm. \textbf{(c)} After $N$ arms have been generated, local complementation (LC) is applied about ancilla $A$, and ancilla $A$ is then measured. The exact sequence of gates for this process is given in the Appendix.}
	\label{fig:generation}
\end{figure*}

Ref.~\cite{Azuma_NC15} presented a family of RGSs where each member of the family consists of $2N$ entangled photons, with $N$ of the photons comprising a fully connected graph at the core, while the remaining $N$ photons are each attached to one of the core photons by a single edge forming $N$ external “arms”. For example, the $N=4$ and $N=6$ RGSs are illustrated in Fig.~\ref{fig:rgs}. In the scheme of Ref.~\cite{Azuma_NC15}, one such state is generated at each primary node of the repeater network, and entanglement swapping between primary nodes is performed by sending half of the $N$ photon arms to an adjacent secondary node, where they encounter an additional $N/2$ photons that were sent out from the repeater state at the next primary node. Entanglement between the primary nodes is then created by performing Bell measurements on pairs of photons that arrived from different primary nodes. The redundancy of sending $N/2$ photons rather than one overcomes the probabilistic nature of the Bell measurements, so that the likelihood of successful entanglement swapping increases with $N$.

Using only the pumping technique developed in Ref.~\cite{Lindner_PRL09}, but carefully choosing when to apply Hadamard gates, large portions of photonic repeater states can be generated using only a single emitter. In particular, the interconnected core photons [see e.g. Fig.~\ref{fig:rgs}] as well as two of the photon arms can be generated in this way using only one emitter. For example, the portion of the $N=4$ RGS that can be created is shown in Fig.~\ref{fig:energylevels}(c). This state can be generated by performing the following sequence of operations on the emitter: ${\cal M}_ZH{\cal P}H{\cal P}H{\cal P}{\cal P}H{\cal P}H{\cal P}H|0\rangle$, followed by single-qubit $X$ and $Z$ gates on the four central photons. In this sequence, $|0\rangle$ denotes the ground state of the emitter, $H$ is a Hadamard gate, $\cal P$ is the pumping operation, and ${\cal M}_Z$ is a final $Z$-measurement performed on the emitter in order to decouple it from the chain of emitted photons. An arbitrarily large interconnected core of $N$ photons can be generated using a similar sequence containing a larger string of pumping operations in the middle: ${\cal M}_ZH{\cal P}H{\cal P}H{\cal P}^{N-2}H{\cal P}H{\cal P}H|0\rangle$.

To date, the most efficient method for constructing RGSs using only probabilistic fusion gates \cite{Pant_PRA17} does so by starting from many three-photon GHZ states (which are themselves generated from parametric down-conversion and fusion) and fusing them together sequentially to build up the necessary entanglement. Constructing the $N=4$ RGS state shown in Fig.~\ref{fig:rgs} in this way would require 5 successful fusion gate applications between pairs of three-photon GHZ states (more generally $2N-3$ fusion gates are needed for an RGS made of $2N$ photons). In contrast, our single-emitter scheme requires only 2 fusion gates to complete the $N=4$ RGS (more generally, $N-2$ fusion gates). Thus, we see that the utilization of a photonic emitter significantly decreases the necessary number of probabilistic fusion gates and the number of photon sources needed at each node, greatly reducing the overhead compared to what is needed to generate the entire state via fusion.

We now show that an additional dramatic reduction in the overhead can be achieved by introducing an ancilla qubit. In particular, we demonstrate that arbitrarily large repeater graph states can be completely generated using only a single emitter coupled to one ancilla qubit, which does not itself need to be an emitter. Inclusion of the ancilla greatly increases the flexibility one has in creating graph states. This is because entanglement between the emitter and the ancilla can be converted into entanglement between the ancilla and the emitted photons using single-qubit gates [see Fig.~\ref{fig:generation}(a)]. The emitter can therefore be used to attach multiple strings of entangled photons to the ancilla. Our scheme for deterministically generating RGSs combines this observation with the fact that RGSs are closely related to tree-like cluster states \cite{Varnava_PRL08}. In particular, the fully interconnected web of photons at the center of an RGS can be obtained by performing an operation known as local complementation (LC)\cite{Bouchet_DM93,Hein_PRA04} around the central ``root'' vertex of a tree, as shown in Fig.~\ref{fig:generation}(c). LC can be implemented by applying the single-qubit gate $e^{i\frac{\pi}{2}\frac{Y+Z}{\sqrt2}}$ to the root vertex and the gate $e^{i\frac{\pi}{2}\frac{X+Y}{\sqrt2}}$ to each of the neighboring vertices. Single-qubit photonic gates are much easier to implement than gates on the emitters; thus, the generation of tree states can easily be extended to repeater states via local complementation.

Fig.~\ref{fig:generation} summarizes our scheme for deterministically generating the all-photonic repeater states introduced in Ref.~\cite{Azuma_NC15}.  Our procedure for generating the underlying tree states involves using the ancilla (labeled $A$) at the root as an ``anchor", and then using the emitter (labeled $B$) to generate each arm of the tree, one at a time. For each arm, the emitter is connected to the root vertex by applying a CZ gate between the emitter and ancilla. The emitter is then pumped twice as shown in the top parts of Figs.~\ref{fig:generation}(a),(b), Hadamard gates are applied to one of the photons and the emitter, and finally a $Z$-measurement is performed on the emitter. This has the effect of severing the emitter from the graph, leaving the other two photons attached to the ancilla in a chain as shown in the bottom parts of Figs.~\ref{fig:generation}(a),(b). The emitter can then be reinitialized, and the whole process can be repeated in order to generate an arbitrary number of arms in the tree cluster state. The full sequence of gates needed to create an RGS with an arbitrary number of  arms $N$ (corresponding to an RGS of degree $N/2$) is given in the Appendix. The most challenging part of the sequence are the emitter-ancilla entangling CZ gates. In the Appendix, we show that the minimum number of CZ gates needed is $N-2$ if the ancilla is also an emitter or $N$ if the ancilla does not emit photons. Thus, if the ancilla is chosen to also be an emitter, the requisite number of two-qubit gates can be slightly reduced.

A crucial feature of our scheme is that all of the photons comprising an arbitrary repeater state can be emitted by the same emitter. This is illustrated in Fig.~\ref{fig:generation}, where the emitter ($B$) produces all of the photons, while the ancilla ($A$) never emits even a single photon. The role of the ancilla is to hold the different arms of the tree together while the emitter generates new arms and attaches them to the ancilla. The fact that all photons are emitted from a single emitter makes it far simpler to ensure that the photons comprising the RGS are indistinguishable, as is necessary for the functionality of repeater networks. It is of course still necessary to achieve uniformity between emitters on different nodes of the network. This will be considered further in the Discussion section.
 
It is also important to emphasize that the role of the ancilla qubit is very different from that of quantum memories in traditional repeater schemes. In traditional schemes, the quantum memories must remain coherent and entangled with photons during the time it takes the photons to reach the secondary nodes, during the time it takes to perform the Bell measurements, and during the time it takes to transmit classical heralding signals between nodes. In contrast, the ancilla qubit in our scheme only needs to be coherent and entangled with photons during the RGS generation process; once the state is formed, the ancilla is no longer required to remain coherent.

\begin{figure}[ht]
	\centering
	\includegraphics[width=\columnwidth]{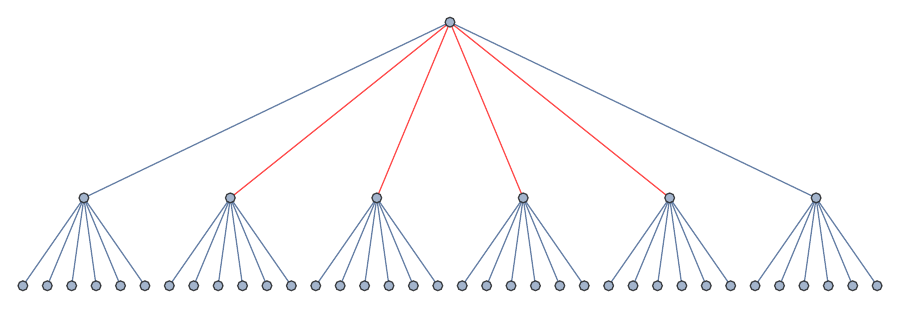}
	\caption{A tree of depth $d=2$ with $k=6$ arms.  Blue edges represent entanglement created by pumping an emitter, while red edges represent entanglement created with a CZ gate between an emitter and another emitter or ancilla.}
	\label{fig:tree}
\end{figure}
Our method can be generalized to deterministically create arbitrary tree states. More general tree states are important because they can be combined with graph states such as the RGS shown in Fig.~\ref{fig:generation}(c) to create a similar state that is robust against photon losses (up to a rate of 50\%) \cite{Varnava_PRL08,Pant_PRA17}. To create a general tree state, one emitter/ancilla is needed for each level of the tree except for the bottommost one. In our proposed scheme, one ancilla is again used at the root as an ``anchor'', and an emitter is used to generate each arm using CZ gates as above. For each subtree, this emitter is then treated as the new ``anchor'' and a second emitter is used to generate each arm of the subtree. This process is continued recursively until the last emitter is pumped repeatedly to create the photons along the bottom of the tree (see Fig.~\ref{fig:tree}). This process only requires CZ gates between emitters/ancillas on neighboring levels, so the scheme can be implemented in any architecture with linearly aligned emitters/ancillas and requires CZ gates only between nearest neighbors. Emitters at every other level can be pumped twice for each vertex at that level, reducing the requisite number of CZ gates by 2 for each instance, as illustrated in Fig.~\ref{fig:tree}. The total number of CZ gates needed to generate a tree of depth $d$ with $k$ arms at each vertex is given by the following formula:
\begin{equation}
N_{CZ}=-1+\frac{k^{d}+(-1)^{d+1}}{k+1}.
\end{equation}
For trees with a large number of arms at each vertex, this is only marginally more efficient; however, this effect can make a drastic difference in the generation of much smaller trees. For a binary tree, in addition to reducing the number of CZ gates by a factor of 3, the number of emitters required is cut in half as well. For a general tree, it is also possible to replace all but one emitter with an ancilla qubit that does not emit photons, albeit at the expense of increasing the necessary number of CZ gates.
\begin{figure}
	\centering
	\includegraphics[width=.45\columnwidth]{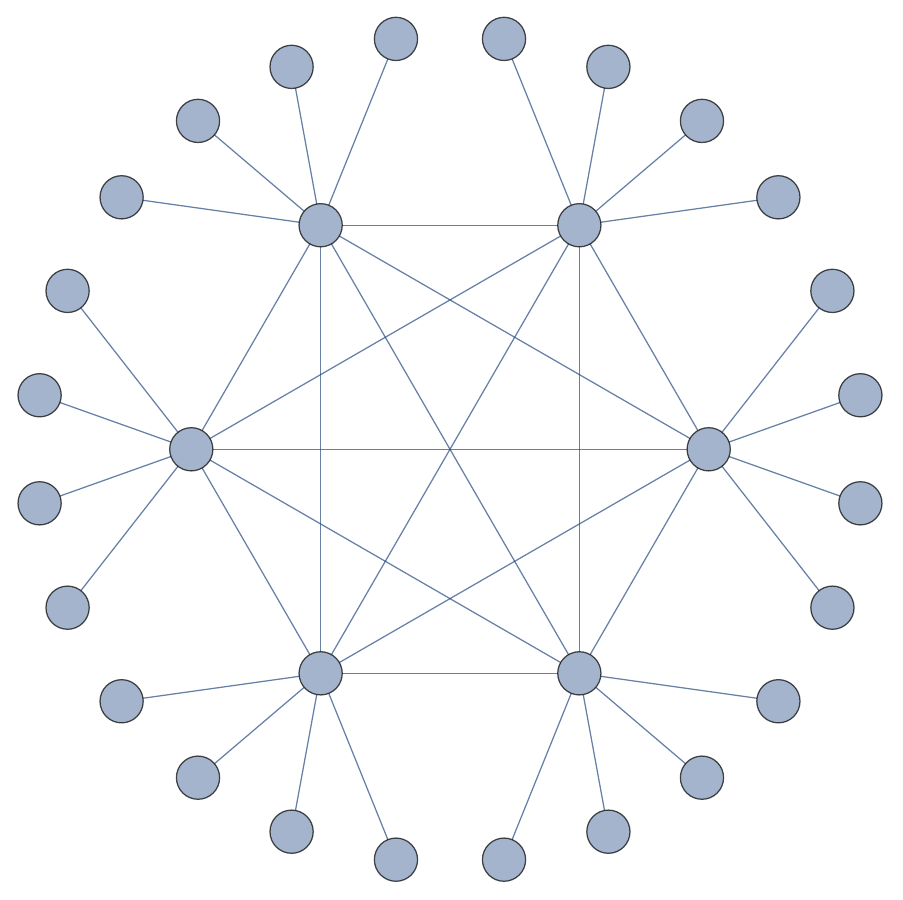}
	\includegraphics[width=.45\columnwidth]{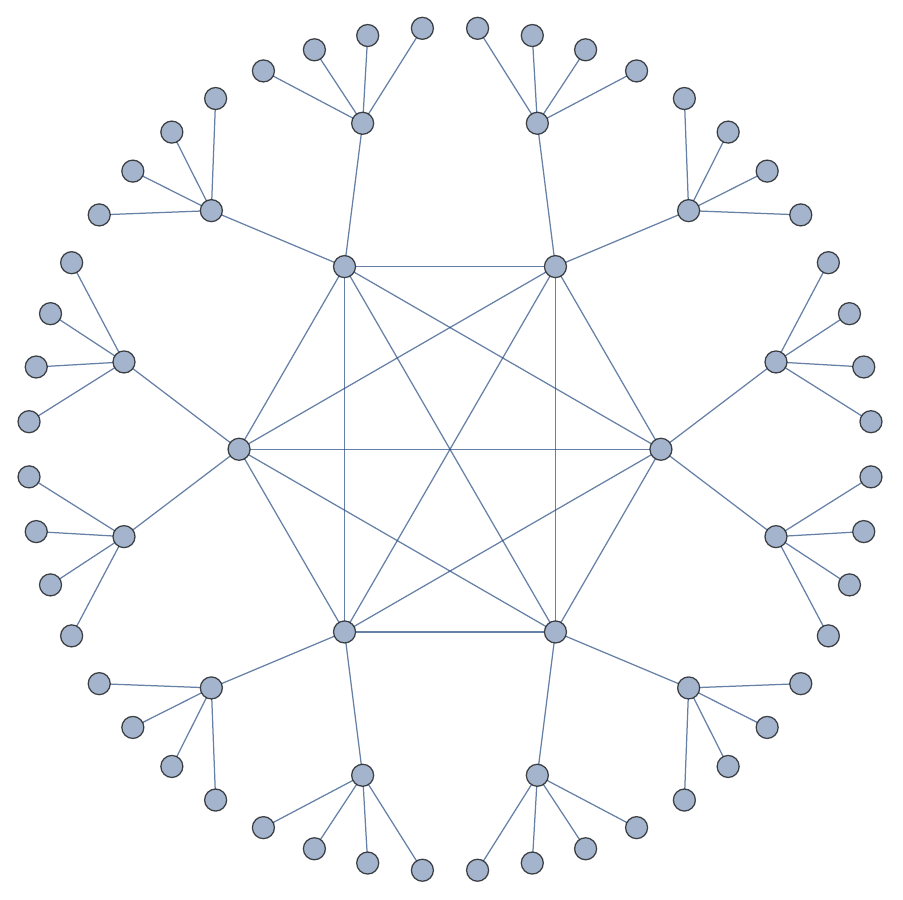}
	\caption{Encoded RGS with (a) depth-one and (b) depth-two tree structures. Both of these can be generated by pumping a single emitter, which is coupled to an ancilla qubit via CZ gates.}
	\label{fig:withtrees}
\end{figure}

\begin{figure}[ht]
	\centering
	\includegraphics[width=\columnwidth]{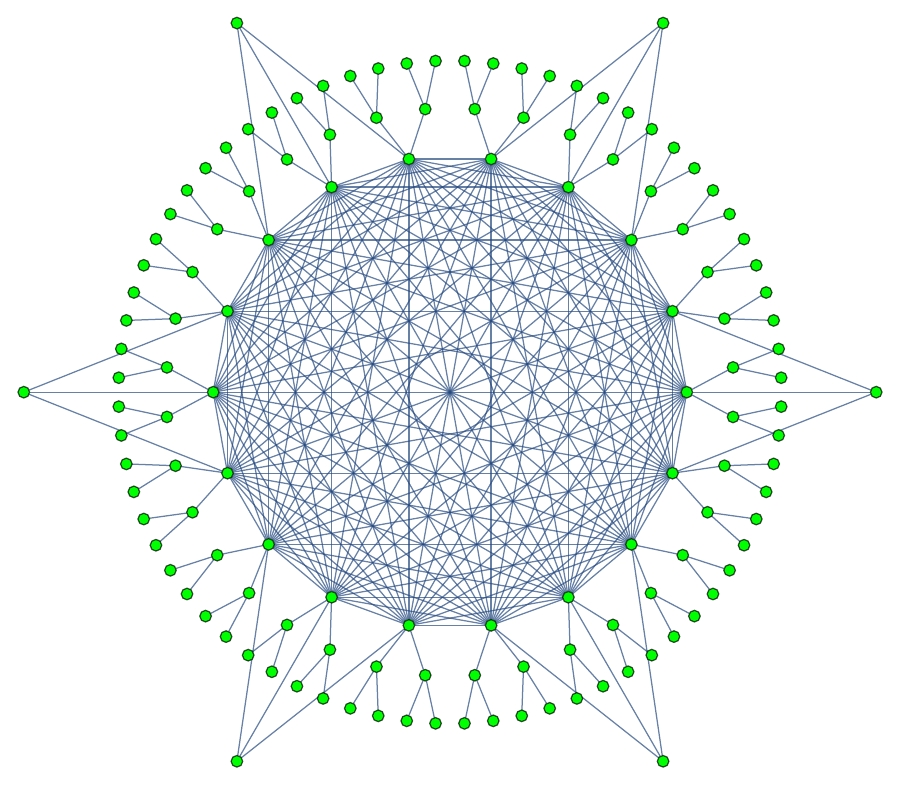}
	\caption{A large repeater state with $N{=}6$ (logical qubits) proposed in Ref.~\cite{Pant_PRA17} which includes subtrees in order to make the state more robust against errors.}
	\label{fig:bbnstate}
\end{figure}

In Ref.~\cite{Pant_PRA17}, it was shown how to combine trees with RGSs to produce a much larger state that has built-in error correction and is robust against photon loss.  The ability to perform $X$ and $Z$ measurements on the central photons is a crucial part of the repeater protocol. For the smaller RGSs discussed above, the loss of one of these central photons would cause the entire process to fail. The addition of trees attached to each of these central photons allows these measurements to be recovered in the case of photon loss \cite{Varnava_PRL06}. Fig.~\ref{fig:withtrees} (a) and (b) show two examples of repeater states with tree structures included; each of these states can be generated from only one emitter and one ancilla qubit. The more complex encoded state proposed in Ref.~\cite{Pant_PRA17}, shown in Fig.~\ref{fig:bbnstate}, can be deterministically generated using our scheme with only two emitters and one ancilla. Creating this state from 3-photon GHZ states using fusion requires roughly one fusion gate per photon, and for the particular size shown, 129 successful fusion gates would be needed. Using two emitters and one ancilla qubit instead, a similar number of pumping operations are needed, but only 24 entangling CZ gates are required between the emitters and ancilla. Alternatively, the emitter which generates the subtrees can be reattached to the middle emitter and pumped to generate the external arms, eliminating the need to pump the second emitter at the cost of $N$ additional CZ gates. For this size, 30 CZ gates are needed, and the process would require two ancilla qubits and only one emitter.

\begin{figure*}[ht]
	\centering
	\includegraphics[width=1.7\columnwidth]{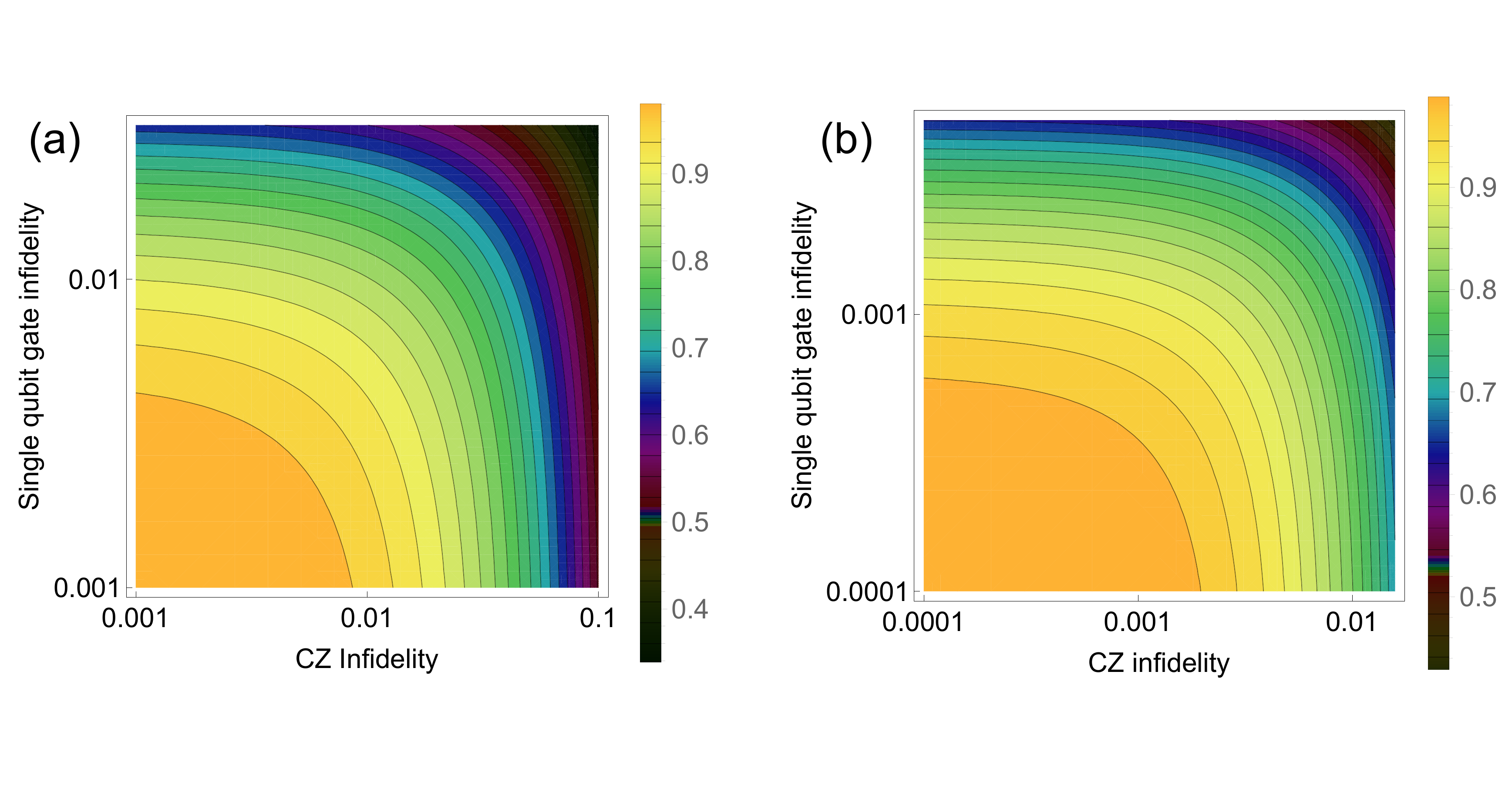}
	\vspace{-1cm}
	\caption{(a) The fidelity of the $N {=} 6$ RGS shown in Fig.~\ref{fig:rgs} as a function of the infidelities of the two-qubit CZ gate and the individual single-qubit gates applied to emitters/ancillas (which are assumed to all have the same infidelity).  (b) The fidelity of the large repeater state shown in Fig.~\ref{fig:bbnstate}.}
	\label{fig:fid1}
\end{figure*}
\begin{figure}
	\centering
	\includegraphics[width=.85\columnwidth]{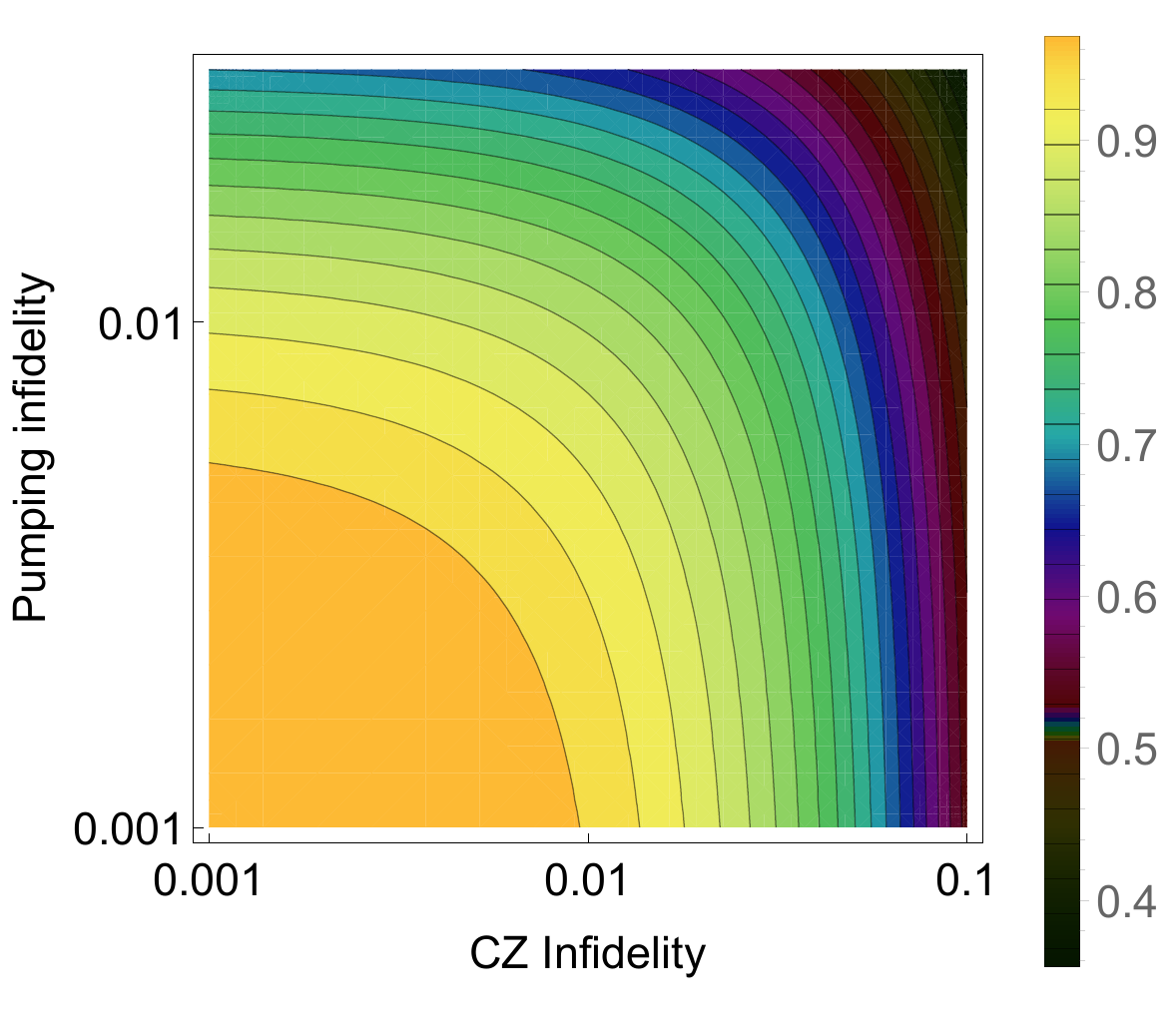}
	\caption{The fidelity of the $N=6$ RGS shown in Fig.~\ref{fig:rgs} as a function of the infidelity of the two-qubit CZ gate and the optical pumping infidelity. (The fidelity of single qubit gates is taken to be 99.9\%.)}
	\label{fig:pumpfid}
\end{figure}
\begin{figure}
	\centering
	\includegraphics[width=\columnwidth]{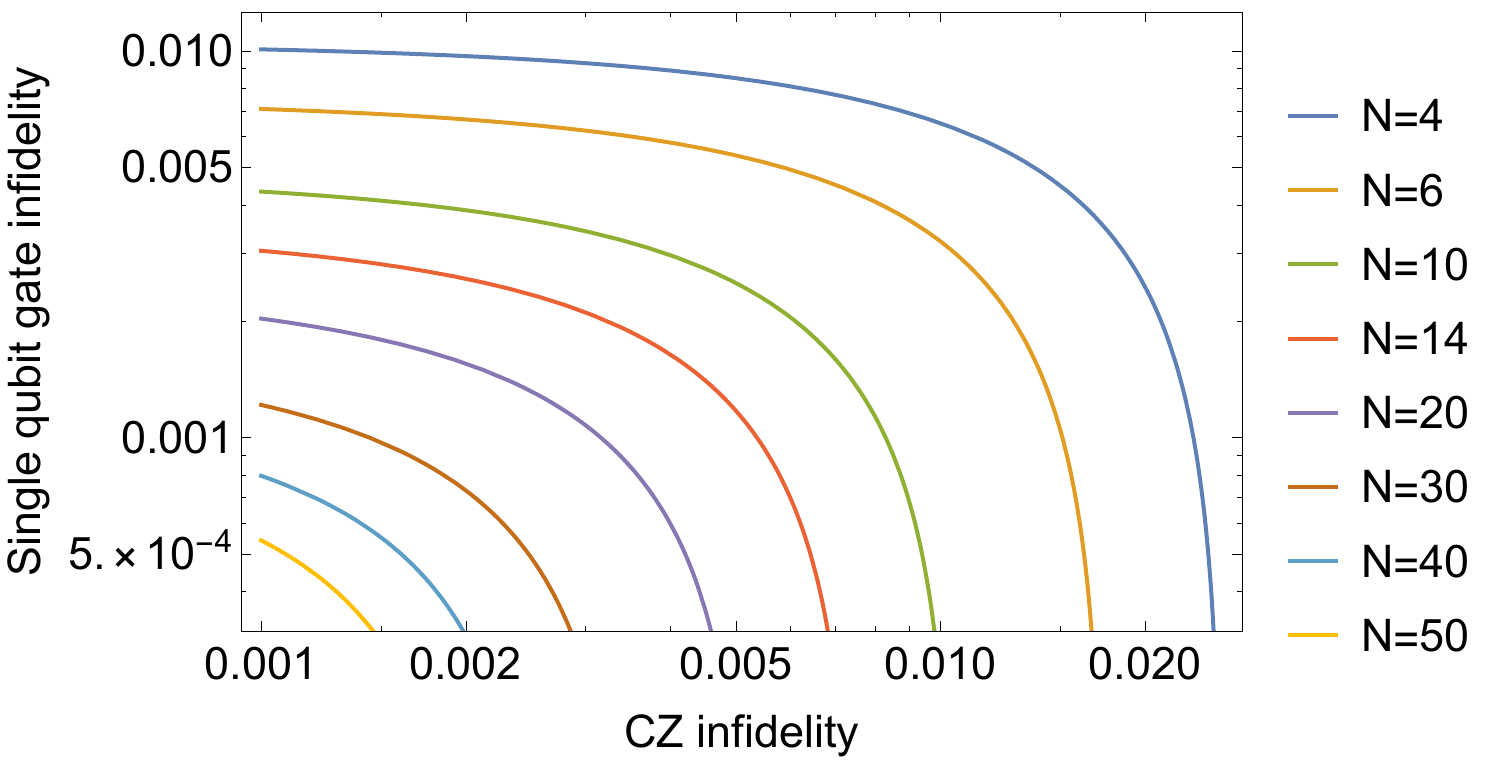}
	\caption{The single-qubit gate and two-qubit CZ gate infidelities needed to create a RGS of size $N$ with 90\% fidelity.  The fidelity of the RGS is given by $CZ^NU^{2N+2}P^{2N}$, where $CZ$, $U$, and $P$ are the fidelities of the CZ gates, single qubit gates on the emitters/ancillas, and pumping operations respectively.}
	\label{fig:fid90}
\end{figure}

Our protocol for generating repeater states has many attractive features. First, RGSs of any size can be generated using only one emitter and one ancilla qubit, which may or may not emit photons. In addition, the only multi-qubit gates required are the CZ gates between the emitter and ancilla. Large trees can be generated with the requisite number of emitters/ancillas scaling as the depth of the tree, thus logarithmically in the total number of photons. We quantify the practicality of our scheme by finding the fidelity of the RGS as a function of the individual gate fidelities of the gates used in the RGS generation sequence. In Fig.~\ref{fig:fid1} we assume that the fidelities of single photonic gates and optical pumping are much higher than the fidelities of single unitary gates and CZ gates on the emitters/ancillas, and thus the fidelity of the final states are essentially determined by the latter two. Fig.~\ref{fig:fid1}(a) shows the fidelity for creating the bare repeater $N=6$ state shown in Fig.~\ref{fig:rgs}, where it is evident that this state can be created with greater than 90\% fidelity if CZ gate fidelities are above 99\% and single-qubit gate fidelities exceed 99.8\%. Fig.~\ref{fig:pumpfid} shows how the RGS fidelity depends on optical pumping, revealing that the demands on the CZ-gate fidelities do not significantly increase for pumping fidelities of around 99.7\%. The way in which these requirements scale with the size of the state is shown in Fig.~\ref{fig:fid90}. We see that increasing the size of the RGS by a factor of 2 requires the infidelity to decrease by roughly half. In the case of the error-correcting repeater state shown in Fig.~\ref{fig:bbnstate}, we see from Fig.~\ref{fig:fid1}(b) that reaching 90\% fidelity requires CZ gate fidelities around 99.8\% and single-qubit gate fidelities around 99.95\%. 

Although the demands on gate fidelity increase with increasing RGS size, it is important to note that modest-sized RGSs may be sufficient for long-distance communication. This is because the probability that at least one successful entanglement swapping operation is achieved between each pair of adjacent primary nodes grows quickly with $N$. In particular, if each Bell measurement succeeds with probability 50\%, then the probability of successfully creating entanglement across the entire network is $(1-2^{-N/2})^{n+1}$, where $n$ is the number of primary nodes in the network. For a network containing $n=1000$ nodes, the probability of success is already 99.9\% using modest-sized RGSs with $N=40$.

One significant advantage of the scheme for generating linear cluster states proposed in Ref.~\cite{Lindner_PRL09} is that it is fault-tolerant, namely that an error which occurs at one point during the process is contained locally and does not propagate to the rest of the state. Generating the photon arms is done using the same method, so an error on emitter $B$ or any of the photons will only affect one specific arm. Thus, for our procedure to be fault-tolerant, it is necessary only to guard ancilla qubit $A$ against errors. In physical implementations, qubit $A$ can be chosen to be a qubit which has a longer coherence time, making errors on qubit $A$ much less likely.  Additionally dynamical decoupling pulses can extend the coherence time of qubit $A$ by orders of magnitude \cite{Maurer_Science12,Muhonen_NatNano14}. Alternatively, our protocol can be modified to make qubit $A$ a logical qubit with built-in error-correction. For example, in physical systems $T_2$ is significantly shorter than $T_1$, which causes $Z$ errors to be the primary errors. In order to guard against these, a logical qubit consisting of 3 physical qubits can be used. This scheme has been demonstrated experimentally using NV center and nuclear spins in diamond \cite{Waldherr_Nature14}. $|0_L\rangle$ is encoded as $|+++\rangle$ and $|1_L\rangle$ is encoded as $|---\rangle$. The logical CZ gate between the emitter $B$ and the physical qubits $1$, $2$, $3$ would be given by $H_1H_2H_3CCZ_{B12}CCZ_{B13}CCZ_{B23}H_1H_2H_3$, where $CCZ$ is a controlled CZ gate. This sequence of gates acts identically on logical states with and without a $Z$ error, meaning that a single $Z$ error could occur at any point during the protocol and no error correction would be required until directly before the final measurement.

\section*{Discussion}

There are several criteria in identifying systems to implement our scheme for deterministic all-photonic repeater state generation. First, the emitters must have the requisite level structure and selection rules, namely they must have two degenerate ground states, each of which is coupled to one corresponding excited state as depicted in Fig.~\ref{fig:energylevels}(a). Note that for the repeater states shown in Figs.~\ref{fig:rgs} and ~\ref{fig:withtrees}, only one emitter needs to satisfy this condition. The second criterion is that emitters/ancillas should be coupled to each other to enable the entangling CZ gates. Third, the emitted photons must be indistinguishable, both at different nodes and at the same node, over the time it takes to generate and entangle repeater states. Fourth, the extraction efficiency of the photons from the emitter should be high. Fifth, the setup should be able to incorporate coupling to fibers in order to transmit the photons to the remote nodes with high fidelity. Finally, it is desirable to have high yield, meaning that the photons are generated at  high rates.

Self-assembled quantum dots (QDs) coupled to cavities satisfy all of these criteria. Their broken symmetry along the growth axis provides the desired level structure and selection rules, and pairs of quantum dots can be grown in a stacked configuration allowing for coupling between the two electron spins trapped in each dot via the exchange interaction \cite{Kim_NatPhys11}. This interaction is the enabling mechanism for the inter-emitter CZ gate (here the ancilla is also a quantum dot emitter).  In fact, the early works for 1D \cite{Lindner_PRL09} and 2D \cite{Economou_PRL10} cluster state generation were based on quantum dots. In a recent breakthrough experiment, Schwartz et al. \cite{Schwartz_Science16} generated deterministically a cluster state string of five entangled photons using a confined dark exciton in a quantum dot. This work constituted a proof-of-principle demonstration that does not yet include optimization over the various metrics (photon generation rate, efficiency, etc). Over the last few years there has been great interest and rapid progress in single-photon devices based on quantum dots which can be harnessed for the repeater generation device we are proposing. One challenge traditionally associated with quantum dots is their spectral inhomogeneity, a severe issue for indistinguishability. Our scheme for bare repeater states avoids this difficulty since all photons can be produced by a single emitter. However, it remains a challenge for photons in different repeater states generated at different nodes. Several years ago, it was demonstrated that quantum dots can be tuned over a large spectral range, which allowed for photon interference coming from remote QDs \cite{Patel_NP10}, and very recently it was shown that such tuning can be successfully implemented in a QD-micropillar device \cite{Ding_PRL16}, demonstrating indistinguishability of photons over more than 10 microseconds \cite{Wang_PRL16}. Another challenge that was recently overcome is the difficulty of photon extraction. Within the last year, several groups have shown photon extraction rates ranging from 66\% \cite{Ding_PRL16} to more than 98\% \cite{Arcari_PRL14}, and very recently the chip-to-fiber coupling efficiency was shown to exceed 80\% \cite{Daveau_preprint16}. These advances stem from enhanced Purcell emission into waveguide modes, simultaneous reduction into other modes, and tapering of waveguides to improve chip-to-fiber coupling. Efforts to engineer these systems to even higher metrics are ongoing, and we anticipate near-ideal metrics over the next couple of years.

Other types of emitters aside from quantum dots can also be used, with recent work showing that several defect centers, including the NV center in diamond and in silicon carbide (SiC), as well as vacancy and divacancy defects in SiC, have the required level structure \cite{Economou_nano16}. Similar challenges as in quantum dots are being addressed for these systems, namely the broad photon emission, coupling to cavities and photon extraction. There has been significant experimental progress recently in terms of emitter indistinguishability \cite{Sipahigil_PRL12}, photon extraction efficiency \cite{Gould_PRApplied16}, and coupling to fibers \cite{Tiecke_Optica15}, and theoretical proposals for quantum networks based on defect centers have been proposed \cite{Childress_PRL06,Nemoto_PRX14,Vinay_PRA17}. Purcell enhancements amounting to a 70-fold increase of emission into the zero phonon line have been achieved in NV centers in diamond \cite{Faraon_PRL12,Li_NC15}, and protocols for suppressing spectral diffusion have been put forward \cite{Fotso_PRL16}. Very recent developments with photonic crystal cavities in SiC have demonstrated an 80-fold enhancement of selective photon emission into the desired zero phonon line, i.e., without enhancing the spectrally closest transition \cite{Bracher_preprint16}. On the other hand, the mechanism for the inter-emitter entangling CZ gate is not as clear as in the case of QDs, where dots can be stacked during growth. Recent experiments, however, have made progress toward controllable defect positioning \cite{Wang_preprint16}; this result can pave the way toward the design and demonstration of an entangling gate between nearest neighbor defects. Since a RGS can be generated from pumping only one emitter, the ancilla qubit could be a different system, such as a nuclear spin \cite{Fuchs_NP11,Dutt_Science07}. 

Our protocol can also be implemented in atomic systems. Trapped ions are particularly promising due to the high level of control that has been demonstrated in these systems and to their ability to be coupled to cavities. To construct a $N=50$ RGS with fidelity 90\%, the required fidelity of single- and two-qubit gates is 99.99\% and 99.85\% respectively (see Fig.~\ref{fig:fid90}). These fidelities have been achieved in trapped ion systems \cite{Ballance_PRL16,Gaebler_PRL16}, for which spin-photon interfaces have also been demonstrated \cite{Blinov_Nature04,Olmschenk_Science09,Stute_Nature12}. Specifically, Ref.~\cite{Stute_Nature12} presented a spin-photon interface with high generation rates ($>$97\%) which moreover has the correct level structure to be compatible with our protocol.

Other systems that show promise are defects or quantum dots in 2D materials, including hexagonal boron nitride and transition metal dichalcogenides, where single photon emitters have been seen \cite{Srivastava_NatureNano15,He_NatureNano15,Koperski_NatureNano15,Chakraborty_NatureNano15,Tran_NatureNano16}. The 2D nature of these materials allows for high photon extraction rates, and coupling to photonic crystal cavities has already been demonstrated \cite{Wu_Nature15}.

With all the recent experimental developments, proof-of-principle demonstrations of our protocols for deterministic repeater graph states can be carried out using existing technology. These demonstrations will likely initially only achieve small sized RGSs due to current limits on two-qubit gate fidelities and to photon generation and extraction efficiencies. Nevertheless, the field of nano-photonics is very active, with higher quality material and device improvements occurring at a rapid pace \cite{Aharonovich_NP16}. These advances will lead to higher fidelities in the implementation of our designs.

An interesting future direction would be a rate-distance analysis of the all-optical repeaters generated with our technique, similarly to what has been carried out for the fusion-based approach \cite{Pant_PRA17}. Another key future effort would be to design in detail physical implementations of our scheme based on, e.g., quantum dots or color centers in solids. Realistic simulations with such systems would provide an important guide to experiment. On-chip architectures using state of the art photonic components could lead to advantages in miniaturization and scalability. 

{\bf Acknowledgments:} This  research  was  supported  by  NSF  (Grant
No. 1741656).

\section*{Appendix: Gate sequences for RGS creation}

Here, we give the explicit sequences of single- and two-qubit gates needed to generate the various repeater or tree-like graph states presented above. The sequence of gates needed to create an RGS of the form introduced by Azuma et al. \cite{Azuma_NC15} with $N$ total arms is
\begin{equation}
\begin{split}
{\cal M}_A(e^{i\frac{\pi}{2}\frac{Y+Z}{\sqrt2}})_A \prod_{n=0}^{N-1}\Big[(e^{i\frac{\pi}{2}\frac{X+Y}{\sqrt2}})_{2n+1}H_{2n+2}{\cal M}_BH_B*\\CZ_{A,B}{\cal P}_{B,2n+2}{\cal P}_{B,2n+1}H_B\Big]*H_A|0\rangle.
\end{split}
\end{equation}
Here, the photons which make up the RGS are labeled 1 through $2N$, $H_n$ represents a Hadamard gate applied to the $n$th photon (the $n$th node of the graph), ${\cal M}_k$ represents a projective $Z$-measurement on emitter/ancilla $k$, $CZ_{A,B}$ is a two-qubit CZ gate acting on ancilla $A$ and emitter $B$, and ${\cal P}_{k,n}$ denotes the pumping/emission of the $n$th photon from emitter $k$. We note that the combination of gates ${\cal M}_A(e^{i\frac{\pi}{2}\frac{Y+Z}{\sqrt2}})_A$ and ${\cal M}_BH_B$ are equivalent to $Y$ and $X$ measurements respectively. CZ gates between the emitter and ancilla are generally the most difficult gates to implement. We can reduce the number of required CZ gates from $N$ to $N-2$ if the ancilla qubit is also an emitter, in which case the sequence becomes
\begin{widetext}
\begin{equation}
\begin{split}
(e^{i\frac{\pi}{2}\frac{X+Y}{\sqrt2}})_{2N-1}H_{2N}{\cal M}_A(e^{i\frac{\pi}{2}\frac{Y+Z}{\sqrt2}})_A{\cal P}_{A,2N}{\cal P}_{A,2N-1}(e^{i\frac{\pi}{2}\frac{Y+Z}{\sqrt2}})_A
\prod_{n=1}^{N-2}\Big[(e^{i\frac{\pi}{2}\frac{X+Y}{\sqrt2}})_{2n+1}H_{2n+2}{\cal M}_BH_BCZ_{A,B}*\\{\cal P}_{B,2n+2}{\cal P}_{B,2n+1}H_B\Big]*
(e^{i\frac{\pi}{2}\frac{X+Y}{\sqrt2}})_1H_2H_A{\cal P}_{A,2}{\cal P}_{A,1}H_A|0\rangle.
\end{split}
\end{equation}
\end{widetext}

$Z$ measurements of emitters can either be performed directly or by pumping the emitter and performing a $Z$-measurement on the photon that is produced. Instead of measuring emitter $B$ between every arm, it is also possible to pump the emitter an extra time and proceed directly to the next arm, postponing the measurements of these photons until after the complete state has been generated. In this case, the sequence becomes
\begin{widetext}
\begin{equation}
\begin{split}
\left(\prod_{n=1}^{N-1}{\cal M}_{3n+2}\right)*(e^{i\frac{\pi}{2}\frac{X+Y}{\sqrt2}})_{3N-3}H_{3N-2}(e^{i\frac{\pi}{2}\frac{X+Y}{\sqrt2}})_{3N-1}{\cal P}_{A,3N-1}(e^{i\frac{\pi}{2}\frac{Y+Z}{\sqrt2}})_A{\cal P}_{A,3N-2}{\cal P}_{A,3N-3}(e^{i\frac{\pi}{2}\frac{Y+Z}{\sqrt2}})_A*\\
\prod_{n=1}^{N-2}\left[(e^{i\frac{\pi}{2}\frac{X+Y}{\sqrt2}})_{3n}H_{3n+1}{\cal P}_{B,3n+2}H_BCZ_{A,B}{\cal P}_{B,3n+1}{\cal P}_{B,3n}H_B\right]*
(e^{i\frac{\pi}{2}\frac{X+Y}{\sqrt2}})_1H_2H_A{\cal P}_{A,2}{\cal P}_{A,1}H_A|0\rangle.
\end{split}
\end{equation}
\end{widetext}

Alternatively any of the $e^{i\frac{\pi}{2}\frac{Y+Z}{\sqrt2}}$, $H$, or $e^{i\frac{\pi}{2}\frac{X+Y}{\sqrt2}}$ gates can be replaced with $e^{i\frac{\pi}{4}X}$,  $e^{i\frac{\pi}{4}Y}$, or $e^{i\frac{\pi}{4}Z}$ respectively. This will produce an equivalent state from which the standard graph state can be recovered by applying $Z$ gates on some of the final photons, depending on which gates were used. Such a correction would also be necessary if emitter $B$ is not reinitialized to $|0\rangle$ between measurement and reentanglement with the rest of the graph state. The particular photons that need to be corrected by $Z$ gates can easily be determined by finding the stabilizer group of the produced state.

\end{document}